\documentclass[twocolumn,showpacs,amsmath,amssymb]{revtex4}

\usepackage{epsfig,psfrag}
\usepackage{dcolumn}% Align table columns on decimal point
\usepackage{bm}% bold math
\usepackage{graphicx}
\usepackage{color}
\newcommand{\beq}{\begin{equation}}
\newcommand{\eeq}{\end{equation}}
\newcommand{\beqr}{\begin{eqnarray}}
\newcommand{\eeqr}{\end{eqnarray}}
\newcommand{\e}{{\epsilon}}

\newcommand{\cb}{{\bar{c}}}

\newcommand{\w}{{\omega}}

\def\bS{{\mathbf S}}

\newcommand{\sigmab}{\mbox{\boldmath $\sigma $}}

\def\cd{c^{\dagger}}
\def\fd{f^{\dagger}}

\def\cb{{\bar{c}}}
\def\fb{{\bar{f}}}
\def\sigmab{{\bar{\sigma}}}
\def\half{{1\over2}}

\def\ua{\uparrow}
\def\da{\downarrow}
\def\eqa{\begin{eqnarray}}
\def\eea{\end{eqnarray}}

\def\ve{{\varepsilon}}
\def\a{{\alpha}}
\def\b{{\beta}}

\def\tj{{\tilde J}}
\def\tjk{{\tilde J_K}}
\def\DKz{{\Delta_{K0}}}
\def\DK{{\Delta_K}}

\def\bcse{\mathbf{S_e}}
\def\bcsf{\mathbf{S_f}}
\def\btau{{\vec{\tau}}}
\def\bh{{\mathbf{h}}}
\def\cs{{\cal{S}}}
\def\eurlet{Europhys. Lett.}
\def\jpa{{Jour. Phys. A}}
\begin{document}

\title{Interplay between the mesoscopic Stoner and Kondo effects in quantum dots}
\author{Ganpathy Murthy} 
\affiliation{Department of Physics and Astronomy,
University of Kentucky, Lexington KY 40506-0055} \date{\today}
\begin{abstract}
We consider electrons confined to a quantum dot interacting
antiferromagnetically with a spin-$\half$ Kondo impurity. The
electrons also interact among themselves ferromagnetically with a
dimensionless coupling $\tj$, where $\tj=1$ denotes the bulk Stoner
transition. We show that as $\tj$ approaches $1$ there is a regime
with enhanced Kondo correlations, followed by one  where the
Kondo effect is destroyed and impurity is spin polarized opposite to
the dot electrons. The most striking signature of the first,
Stoner-enhanced Kondo regime, is that a Zeeman field increases the
Kondo scale, in contrast to the case for noninteracting dot
electrons. Implications for experiments are discussed.

\end{abstract}
\vskip 1cm 
\maketitle

In the simplest version of the Kondo effect\cite{kondo}, a
spin-$\half$ magnetic impurity interacting antiferromagnetically
(exchange coupling $J_K$) with delocalized conduction electrons forms
a singlet with a cloud of conduction electrons. The nonperturbative
Kondo energy scale $\Delta_K\simeq D\exp{-1/J_K\rho_0}$ (where
$\rho_0$ is the density of states per spin per unit volume and $D$ is
a high-energy cutoff) characterizes a host of properties of the
system, including the reduction in ground state energy, the
temperature dependence of the magnetic susceptibility, etc. Recent
advances in nanofabrication have made new mesoscopic realizations of
the Kondo problem possible. In one such
realization\cite{kondo-meso-theory,kondo-meso-expt}, a small quantum
dot with an odd number of electrons in a Coulomb Blockade valley plays
the role of the impurity spin, while the conduction electrons live in
the leads.

In this paper we consider a slightly modified setup in which the
``conduction'' electrons live in a large quantum dot with level
spacing $\delta$, a variant of which has been
realized\cite{marcus-triple-dot}. Such a model with {\it
noninteracting} conduction electrons has been considered
before\cite{kondo-box,kondo-box-signatures}, as has a model where the
impurity spin interacts with a Luttinger liquid\cite{kondo-luttinger},
but hitherto a treatment of realistic interactions between the
``conduction'' electrons in a quantum dot with the Kondo effect has
been lacking.

Recently progress has been made in characterizing interactions in
disordered quantum dots\cite{qd-reviews}, where the Thouless energy
$E_T=\hbar/\tau_{erg}$ plays an important role ($\tau_{erg}$ is the
time it takes for an electron to ergodicize over the dot).  In the
limit where the Thouless number $g=E_T/\delta$ becomes large the
following ``Universal Hamiltonian'' has been proposed for
time-reversal invariant systems\cite{H_U,univ-ham}.
\beq
H_U=\sum\limits_{\a,s}\e_{\a}c^{\dagger}_{\a,s}c_{\a,s}+{U_0\over
2}{\hat N}^2 -J\bS^2+\lambda T^{\dagger} T
\label{univ-ham}\eeq 
Here ${\hat N}$ is the total particle number, $\bS$ is the conserved
total spin, and $T=\sum c_{\b,\da}c_{\b,\ua}$. In addition to the
charging energy, $H_U$ has an exchange energy $J$ and a
superconducting coupling $\lambda$. For semiconductor quantum dots
with $r_s\simeq1$, $J$ is estimated\cite{univ-ham} to be $0.3\delta$,
while $\lambda$ is negligible. Using the fermionic renormalization
group\cite{shankar}, this Hamiltonian has been shown to be a stable
fixed point at weak coupling (small $r_s$), with other phases possible
at strong coupling\cite{qd-us}. We will use $H_U$ as prescribing the 
interactions among the electrons in the dot.  As $J$ becomes stronger
the system undergoes transitions\cite{H_U,univ-ham} to higher and
higher total spin $S$, until at $J=\delta$ the dot becomes
macroscopically polarized in a Stoner transition. Mesoscopic
(sample-to-sample) fluctuations of the magnetization at a given $J$
due to variations of the energy levels have been theoretically
characterized\cite{H_U,univ-ham} and observed\cite{dot-spin-measure}.

The focus of this paper is the interplay between the Kondo and
mesoscopic Stoner effects. Define $\tj=J/\delta$, $\tjk=J_K/\delta$,
and $E_S^0=S^0\delta$, where $S^0$ is the total spin of the dot in the
absence of the Kondo coupling, and denote $\DKz$ as the Kondo scale
for $\tj=0$. Our central result is that in the limit when $g$, $S^0$,
$\DKz/\delta$ are large there are two regimes. In Regime I,
$E_S^0\le2\DKz$, the total spin $S$ is suppressed below $S^0$, while
the Kondo scale $\DK$ is enhanced over $\DKz$. The most significant
signature is that a Zeeman field {\it increases} $\DK$ in this
regime. In Regime II, the Kondo effect is destroyed by the mesoscopic
Stoner effect, and the impurity is almost fully polarized opposite to
the dot spin. There is a large jump in $S$ at the transition in our
mean-field analysis, though a more accurate analysis may reveal a
smooth crossover rather than a transition.

Our model Hamiltonian for a closed dot (not connected to leads)
interacting with an impurity spin, ignoring the Coulomb term (for a
constant number of particles), is
\beq
H=\sum\limits_{ks} \ve_k \cd_{ks}c_{ks} -\tj \delta\bS^2 +{\tjk\delta\over2}\bcsf
\cdot\sum\limits_{kk'ss'}\cd_{ks}\btau_{ss'}c_{k's'}
\eeq
Here $\bcsf$ is the impurity spin and $\btau$ are the Pauli spin
matrices. Since we are interested in the effects of competing
interaction terms and not in mesoscopic fluctuations, we will make the
level spacing uniform ($\ve_k=(k+\half)\delta$) with equal couplings
to all levels. The high energy cutoff is $D=\ve_M$, and thus $k$ goes
between $-M-1$ and $M$. The local electronic spin at the impurity site
$\bcse=\sum\limits_{kk'ss'}\cd_{ks}\btau_{ss'}c_{k's'}/4M$, the
impurity spin $\bcsf$, and the total electronic spin $\bS$ are {\it
not} conserved, but the total spin of the system,
$\bS_{tot}=\bS+\bcsf$ is conserved due to the spin-rotational
invariance of $H$, as is $S_{tot}^z$. As $\tj$ increases, we expect
transitions to successively higher values of $S_{tot}$, exactly as in
the mesoscopic Stoner effect\cite{H_U,univ-ham}.  If the dot has an
even number of electrons, $S_{tot}=p+\half$. We will work in the state
$S_{tot}^z=S_{tot}$.

There are many ways to analyze the Kondo problem\cite{hewson}, with
simplest way for our purposes being the large-$N$
approximation\cite{chak-largeN,read-newns}. In this approach, one
writes the impurity spin in terms of an $f$-electron $\bcsf=\half
\fd_s\btau_{ss'} f_{s'}$, and extends the spin to a degeneracy quantum
number $m$, with $-N/2\le m\le N/2$. To represent the impurity spin
properly a constraint on the number of $f$-electrons is imposed
($n_f=N/2$\cite{chak-largeN} or $n_f=1$\cite{read-newns} which are
identical for $N=2$). Despite some subtle issues concerning the
restoration of symmetry by quantum fluctuations\cite{read2} the
leading large-$N$ results give a fairly good nonperturbative
description of the physics, and are consistent with the results
obtained by other methods\cite{hewson}. We will take the leading
large-$N$ approximation literally for $N=2$, which should capture the
physics of interest. One decouples the Kondo interaction by a
Hubbard-Stratanovich transformation\cite{chak-largeN,read-newns}. We
will also decouple the Stoner interaction $-J\bS^2$ by a
Hubbard-Stratanovich transformation to get
\beqr
Z=&\int D\bh D\cb Dc D\fb Df D\sigmab D\sigma D\lambda e^{-\cs}\nonumber\\
\cs=&\int dt \bigg({\bh^2\over 4J}+2\delta{\sigmab\sigma\over\tjk}+\sum\limits_{s}\fb_s(\partial_t+\ve_f+i\lambda)f_s\nonumber\\
&+\sum\limits_{kss'} \cb_{ks}((\partial_t+\ve_k)\delta_{ss'}-{\bh\over2}\cdot\btau_{ss'})c_{ks'}\nonumber\\
&-in_f\lambda+\sigma\delta\sum\limits_{ks}\cb_{ks}f_s+\sigmab\delta\sum\limits_{ks}\fb_sc_{ks}\bigg)
\eeqr
where the field $\lambda$ imposes the constraint.  At the mean-field
level this describes a set of electrons in the quantum dot hybridizing
with the impurity site and subject to a Zeeman field $\bh$. The
fermionic part of the action can be integrated out to yield the
effective action, the parameters of which must be chosen to lie at a
saddle point, and to satisfy the
constraint\cite{chak-largeN,read-newns}. The saddle-point values of
$\ve_f$ and $\lambda$ are zero for our case. In our maximally
polarized state, $h_z$ has an expectation value, while $h_{x,y}$ are
fluctuating. In order to obtain the values of $\tj$ where the total
spin changes, we will need to keep terms of order $p^2$ and terms of
order $p$ in the effective action.  The analysis is particularly
simple in the limit $D\to\infty$, $\tjk\to 0$, with $\DKz$ held
fixed. In this limit, defining $b=h_z/2$ and $\DK=|\sigma|^2\delta$,
the $j^{th}$ root of the single-particle Green's function is
\beq
\w_{j\ua}=j\delta-b-{\delta\over\pi}\tan^{-1}((j\delta-b)/\DK)
\label{roots}\eeq
with $b\to -b$ for the $\da$ spin. The errors in this are of order
$\delta/\DK$, and can be neglected in Regime I, $\delta\ll
E_S^0\le\DKz$.  The ground state has $-M\le j\le p$ filled for the
$\ua$ spin, while states $-M\le j\le -p-1$ are filled for the $\da$
spin.

In the limit of $T\to 0$ the fermionic contribution to the effective
action is the ground state energy, and we obtain for the static
mean-field effective action at $S_{tot}=p+\half$
\beqr
{\cs_{eff}^{MF}\over\beta}=&{b^2\over \tj\delta}+\delta p(p+1)-{2\over\pi}(b-E_S)\tan^{-1}{{b-E_S\over\DK}}\nonumber\\
&-2bS_{tot}+{\DK\over\pi}\bigg(\log{\bigg({(E_S-b)^2+\DK^2\over\DKz^2}\bigg)}-2\bigg)
\eeqr
where we have used $\DKz=D\exp{-1/\tjk}$ to eliminate references to
$\tjk$. It can be seen that the minimum of $b$ will be close to $\tj
E_S$. We still do not have all the terms in the effective action to
order $p$, for which we have to address fluctuations in $h_{x,y}$.
The action for $h_{x,y}$ is expressed in terms of the suscpetibilities
of the spin operators $S_{x,y}$. Since there is an average $S_z$,
$h_x$ and $h_y$ will have cross terms (a consequence of
$[S_x,S_y]=iS_z$).  Fluctuations of $h_{x,y}$ of higher order than
quadratic are suppressed by powers of $1/b$, which correspond to
powers of $1/S_{tot}$ and can therefore be ignored for large $S_{tot}$
near $\tj\approx 1$. Integrating out the quadratic fluctuations leads
to an effective action correct up to terms of order $p$, which is
\beqr &{\cs_{eff}\over\beta}={b^2\over \tj\delta}+\delta
p(p+1)-2bS_{tot}-{2\over\pi}(b-E_S)\tan^{-1}{{b-E_S\over\DK}}\nonumber\\
&+{\DK\over\pi}\bigg(\log{{(E_S-b)^2+\DK^2\over\DKz^2}}-2\bigg)-b+b|1-2\tj\delta
b F| 
\label{seff-final0}\eeqr
where  the sum $F(p,b,\DK)$
\beq
F=\DK^2\sum\limits_{-p}^{p} {(\tan^{-1}{b+m\delta\over\DK}+\tan^{-1}{b-m\delta\over\DK})^2\over(\DK^2+(b+m\delta)^2)(\DK^2+(b-m\delta)^2)}
\eeq
arises from ``diagonal'' excitations $j\ua\to j\da$ which dominate the
susceptibilities.  For $E_S\ll\DK$, noting that the saddle
point value of $b$ is very close to $E_S$, we  get
\beq
F\approx {S_{tot}\over2b^2}(1-{4b^2\over3\DK^2}+\cdots)
\label{f}\eeq
In this regime, after ignoring the $(b-E_S)$ term which is negligible,
the effective action takes the form
\beqr 
{\cs_{eff}\over\beta}=&b^2({1\over \tj \delta}+{4\tj\delta
S_{tot}\over3 \DK^2})+\delta p(p+1)- 2bS_{tot}-JS_{tot} \nonumber\\
&+{\DK\over\pi}\bigg(\log{{(E_S-b)^2+\DK^2\over\DKz^2}}-2 \bigg)
\label{seff-final1}\eeqr
The additional $b^2$ term in Eq. (\ref{seff-final1}) suppresses
$E_S$. Also, $S_{tot}\ne 0$ favors a larger $\DK$. This term arises
from a smaller gain in spin fluctuation energy at larger $E_S/\DK$
(Eq. (\ref{f})). For $E_S\ll\DK$, since the impurity spin is locked
into a singlet, the entire spin $S_{tot}$ is carried by the dot
electrons, with the corresponding energy gain
$-JS_{tot}(S_{tot}+1)$. With increasing $E_S/\DK$ the spin is
distributed between the dot electrons and the impurity spin, leading
to a smaller gain in $-J\bS^2$. {\it Since transitions between states
of different $S_{tot}$ are driven by the delicate balance\cite{H_U}
between the increase in kinetic energy and gain in spin exchange
energy, this physics is central to the interplay between the
Kondo and mesoscopic Stoner effects}.

\begin{figure}[h]
%\narrowtext
%\epsfxsize=2.4in\epsfysize=2.4in
\includegraphics*[width=2.4in,angle=0]{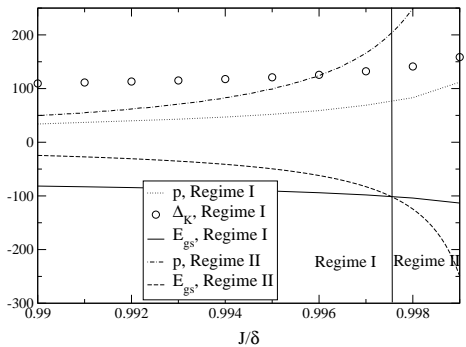}
%\vskip 0.15in
\caption{The variation of $p$, $\DK$, and the ground state energy with
$\tj$ for $\DKz=100\delta$. The solid line represents the ground state
energy of Regime I, while the dashed line represents that of Regime
II. Their crossing, demarcated by the solid vertical line, is the
transition. Note that $\DK$ continuously increases in Regime I, and
that there is a large jump in $p$ at the transition. }
\label{fig1}
\end{figure}

 In Fig. \ref{fig1} we show the result
of a numerical calculation of the minimum of Eq. (\ref{seff-final0})
for $\DKz=100\delta$ as a function of $\tj$. The enhancement of $\DK$
over $\DKz$, and the suppression of $E_S$ below $E_S^0$ are evident
throughout.

\begin{figure}[ht]
%\narrowtext
%\epsfxsize=2.4in\epsfysize=2.4in
\includegraphics*[width=2.4in,angle=0]{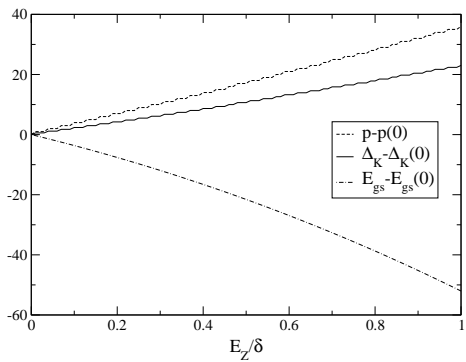}
%\vskip 0.15in
\caption{The variation of $p$, $\DK$, and the ground state energy with
$E_Z$ for $\DKz=100\delta$ and $\tj=0.99$. Their values at $E_Z=0$
have been subtracted out, and are $p(0)=34$, $\DK(0)=109.5$, and
$E_{gs}(0)=-81.67$. Note that $E_Z$ is in units of $\delta$, and even
for $E_Z=0.5\delta$ there is a 10\% enhancement of $\DK$.}
\label{fig2}
\end{figure}

The clearest signature of this state lies in its response to a Zeeman
field $E_Z$, which adds the term $-E_ZS_{tot}$ to
Eq. (\ref{seff-final0}). This term favors larger $S_{tot}$, which as
we have seen in the previous paragraph, favors larger $\DK$. This
effect is displayed in Fig. \ref{fig2} for $\DKz=100\delta$, and
$\tj=0.99$. This is to be contrasted with the usual paramagnetic Kondo
state, in which a Zeeman coupling suppresses $\DK$. This signature can
be seen experimentally as enhancement of the Kondo resonance in the
conductance provided the large dot is weakly coupled to
leads\cite{kondo-box-signatures}. It may already have been
seen\cite{marcus-triple-dot}, about which more below.

Let us now turn to the competing state. For $S_{tot}=p+\half$, the
allowed values of $S$ are $p$ and $p+1$. Taking into account only the
ground state configurations of the dot electrons, one finds the fully
polarized state to be
\beqr
|\Omega\rangle=&\bigg({2p+2\over2p+3}\bigg)\bigg(|p+1,p+1\rangle_e\otimes|\half,-\half\rangle_f\nonumber\\
&-{1\over\sqrt{2(p+1)}}|p+1,p\rangle_e\otimes|\half,\half\rangle_f\bigg)
\eeqr 
which has the impurity spin almost fully polarized opposite to the dot
spin. Consequently, there will be no Kondo coherence in this
state. The coupling to the state with $S=p$ is negligible in the
large-$p$ limit. The energy of this state is
\beq
E_{II}=\delta(p+1)^2-\tj\delta(p+1)(p+2)-\delta{\tjk\over2}(p+2)
\label{EII}
\eeq
There are perturbative corrections to this state and its energy coming
from particle-hole excitations (which can be taken into account as a
{\it static} impurity problem since the impurity spin is effectively
frozen), but the scale of these corrections can be shown to be $\tjk
E_S$. In the limit $D\to\infty$, $\tjk\to0$, keeping $\DKz$ fixed, the
last term of Eq. (\ref{EII}) and the perturbative corrections are
negligible. Fig. \ref{fig1} also shows the energy of this state (in
the above limit) and its total spin. The vertical line denotes the
first-order transition (which may be smoothed to a crossover in a
more accurate calculation) at which there is a large jump in the
spin. This transition is located at roughly $E_S^0\approx 2\DKz$ in
our mean field model. The lowest collective excited state in Regime II
flips the impurity spin, with an energy of order $\tjk E_S$, and
should appear as a resonance in the conductance.

Let us now tie up some loose ends. Our mean-field approximation gives
an accurate picture of the electronic spectrum deep in Regime I, and
there the physics described after Eq. (\ref{seff-final1}) is
robust. However, close to the mean-field transition, our approximation
may be inaccurate, and the transition found in mean-field may be
smoothed into a crossover. We have considered an even number of dot
electrons. For large $S_{tot}$ there is no qualitative difference
between even and odd numbers of dot electrons. While we have carried
out the calculation with $E_S^0,\DKz\gg\delta$, no qualitative
difference is expected with $E_S^0,\DK$ are a few times
$\delta$. There should still be a sharp crossover from a regime with
Kondo coherence to one without as $\tj$ increases. Finally, we have
assumed equal spacings and couplings to the Kondo spin, whereas in
reality both of these are controlled by Random Matrix
Theory\cite{qd-reviews}. For large $\DKz$ and $\tj$ close to $1$, the
Kondo part of the physics is much the same\cite{chak-nayak}. The main
change will be that there are large mesoscopic fluctuations of $S$
(Kurland {\it et al} in ref. \cite{H_U}), which may result in large
mesoscopic fluctuations of the transition point. For smaller $\DKz$,
$E_S^0$, a numerical calculation along the lines of
refs. \cite{H_U,rmt-kondo} needs to be carried out.

Consider now the experimental signatures. Kondo correlations, which
can be seen by their conductance signatures\cite{kondo-box-signatures}
when the dot is weakly coupled to leads, are present in Regime I, and
absent in Regime II. The total spin of the state can be measured by
tracking the movement of conductance peaks as a function of parallel
magnetic field $B_{\|}$\cite{dot-spin-measure}, and a large change
should be seen in the total spin at the
transition/crossover. Experimentally, the clearest signature is the
strong enhancement of $\DK$ with the Zeeman coupling $E_Z$ in Regime
I, as shown in Fig.\ref{fig2}.  Since the spin of the Regime II state
is much larger, a large enough Zeeman coupling will eventually push
the system over into Regime II. In a recent experiment on a system
with a large dot and two smaller dots serving as the impurity
spins\cite{marcus-triple-dot}, the authors observe that the zero-bias
Kondo peak grows stronger with $B_{\|}$ (a signature of Regime I)
before disappearing at zero bias by splitting into two peaks at finite
bias (possibly the $\pm\tjk E_S$ resonances of Regime II). In that
experiment\cite{marcus-triple-dot} both the dots were strongly coupled
to leads, so these observations are suggestive (but not conclusive)
evidence for our physical picture.

In summary, we have analyzed a model in which electrons on a large
quantum dot interact with themselves with a Universal Hamiltonian
ferromagnetic exchange, while also interacting antiferromagnetically
with an impurity spin. We find two regimes: In Regime I (the
Stoner-enhanced Kondo regime) there is a robust Kondo scale which is
enhanced as either $\tj$ or the Zeeman coupling increases. For large
enough $\tj$ or $E_Z$ the system will make a transition/crossover into
Regime II, in which the Kondo coherence is destroyed in favor of
polarizing the impurity spin opposite to the total spin of the dot
electrons. There is a large change in the total spin at the
transition/crossover. These are both mesoscopic regimes, in which the
magnetization per particle can be made as small as one wishes.

There are a number of directions in which this work can be extended,
the most natural and experimentally relevant being the investigation
of mesoscopic fluctuations\cite{H_U,rmt-kondo}. Another theoretically
interesting system is the two-impurity Kondo
problem\cite{two-impurity-kondo}, which for noninteracting conduction
electrons has an unstable non-fermi-liquid critical point. Since such
a triple-dot system has already been realized
experimentally\cite{marcus-triple-dot}, the study of the effects of
the Universal Hamiltonian exchange on two-impurity Kondo physics would
be very interesting and timely. Finally, the case when the dots are
strongly coupled to the leads demands closer scrutiny.

It is a pleasure to thank Assa Auerbach for stimulating discussions
and the NSF for partial support under DMR-0311761.

\end{document}